\documentclass[a4paper]{jpconf}
\usepackage{graphicx}
\usepackage[utf8]{inputenc}
\usepackage{subfigure}
\newcommand{\dd}{\mathrm{d}}

\begin{document}
\title{Additional strange resonances from Lattice QCD}

\author{Michał Marczenko}
\address{Institute of Theoretical Physics, University of Wrocław, pl. Maksa Borna 9, PL-50204 Wrocław, Poland}
\ead{michal.marczenko@ift.uni.wroc.pl}

\begin{abstract}
	Recent Lattice QCD (LQCD) studies suggest that there are missing resonances in the strange sector of the Hadron Resonance Gas (HRG) model. By adopting the continuous Hagedorn mass spectrum, we present how different medium compositions influence the HRG predictions of conserved charge fluctuations. It is shown that missing strange resonances may be partially accounted for by applying the Hagedorn mass spectrum extracted from experimentally established hadrons. On the other hand, the strange-baryonic spectra, extracted from LQCD results for fluctuations, are found to be consistent with the unconfirmed states in the Particle Data Group (PDG) database, whilst the strange-mesonic spectrum points towards yet-undiscovered states in the intermediate mass region.
\end{abstract}

\section{Introduction}
\label{intro}

	The confined phase of QCD, the theory of strong interactions, is commonly modeled with the Hadron Resonance Gas (HRG)~\cite{BraunMunzinger:2003zd}. Its predictions of the QCD equation of state at finite temperature were confirmed by Lattice QCD (LQCD) findings~\cite{Borsanyi:2013bia, Bazavov:2012jq, Borsanyi:2011sw}. Recent LQCD results for fluctuations of conserved charges, however, have pointed to limitations in the HRG description~\cite{missing}. In particular, HRG results for the baryon-strangeness correlations and strangeness fluctuations are below those from LQCD (Fig.~\ref{fig:fluct}), whilst the results for the equation of state and baryon number fluctuations remain in agreement. The discrepancy in the \mbox{strange-hadronic} sector of HRG was attributed to missing resonances in the Particle Data Group (PDG) database~\cite{pdg}.

	To quantify the LQCD equation of state in the confined phase, many extensions of HRG were proposed. One such extension, which gained a lot of attention over the years due to its simplicity and pertinence in phenomenology of particle production, assumes an asymptotic continuous exponential mass spectrum. It was introduced by Rolf Hagedorn in the context of the Statistical Bootstrap Model~\cite{Hagedorn:1965st, Hagedorn:1971mc} and then explored within many others.

	 The experimental PDG spectrum, unlike the Hagedorn one, saturates \mbox{at $\sim 2.5$ GeV}. This may be due to the complicated decay properties and large widths of the heavy resonances, which make their final assessment and experimental confirmation challenging. Hence, the Hagedorn mass spectrum, because of its asymptotic behavior, accounts for possible undiscovered heavy resonances.

	In this contribution, we study whether the inclusion of the heavy resonances, by adopting the Hagedorn mass spectrum, can reduce or resolve the disparities between the HRG and LQCD results on baryon-strangeness and strangeness fluctuations. In order to identify the possible origins of the missing resonances, we study the influence of different medium compositions on  fluctuations of conserved charges. In particular, we compare the predictions of the HRG model and its extended version with the continuous Hagedorn mass spectrum, extracted from the experimentally established discrete PDG spectra and LQCD results on the baryon-strangeness correlations and strangeness fluctuations.

\section{Hadron Resonance Gas and Hagedorn mass spectrum}
\label{section:HRG}

	\begin{figure*}[t!]
		\centering\subfigure[]{\includegraphics[width=.45\columnwidth]{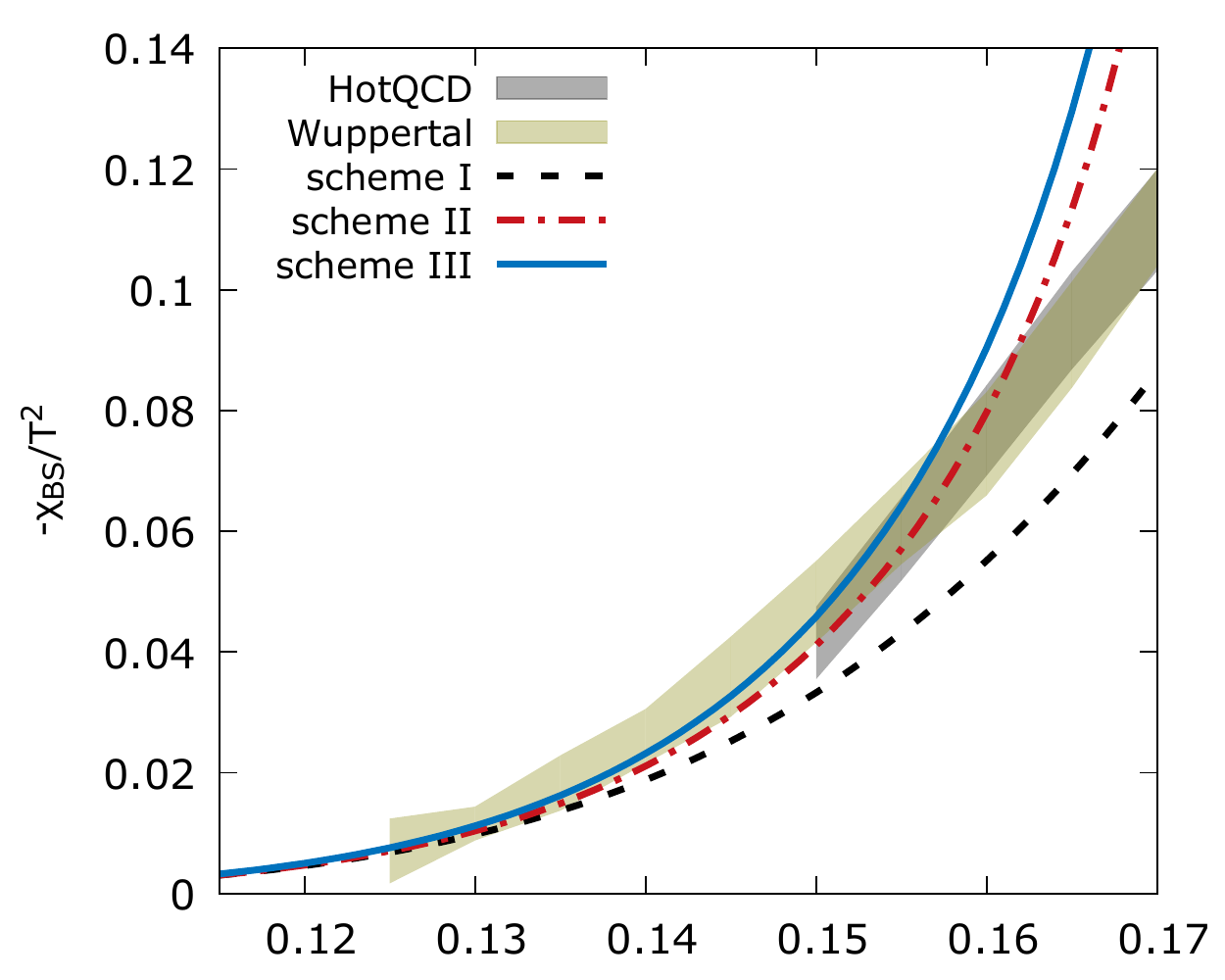}
			\label{fig:fluct:bs}}
		\centering\subfigure[]{\includegraphics[width=.45\columnwidth]{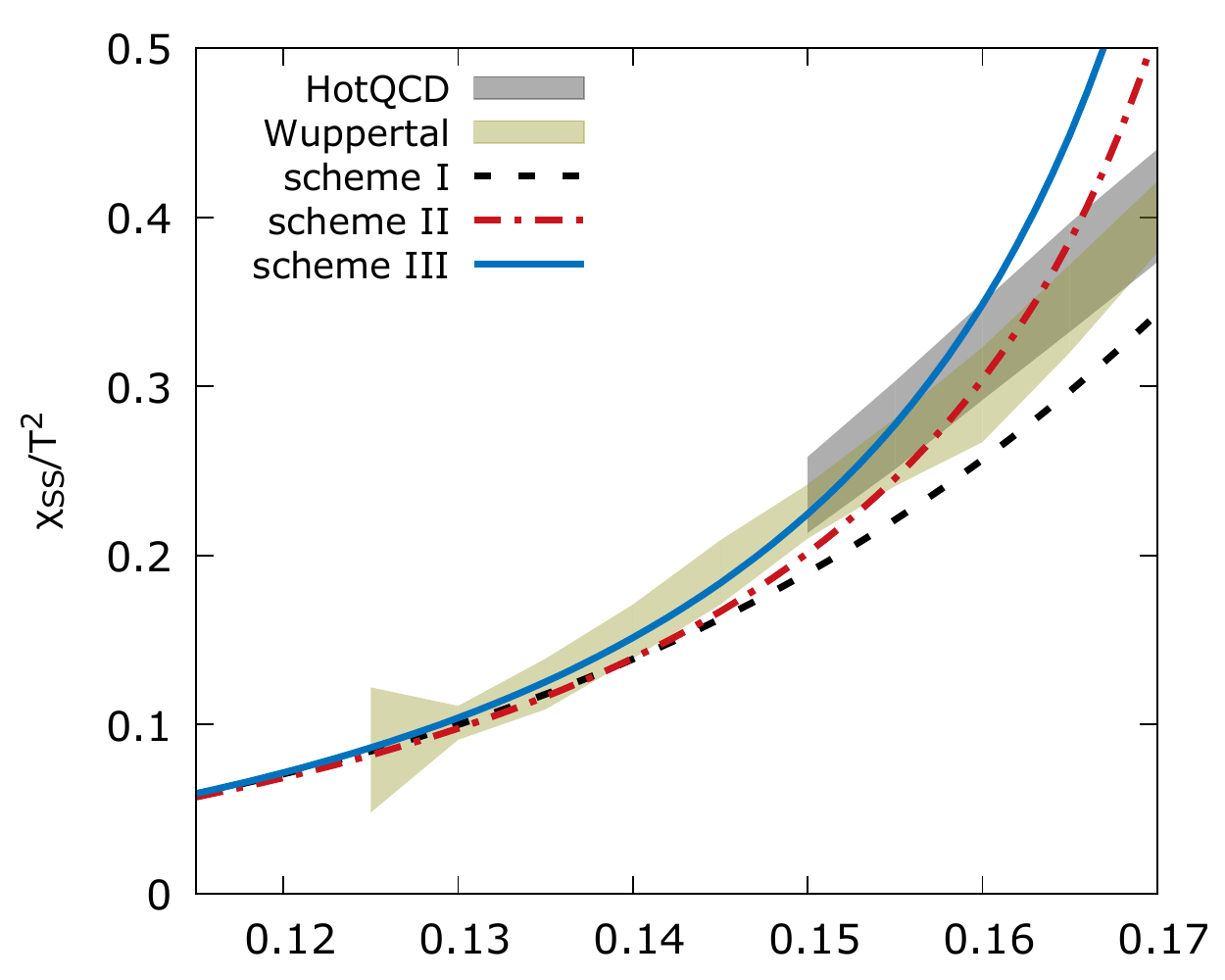}
			\label{fig:fluct:ss}}
		\caption{Lattice QCD results of HotQCD~\cite{Bazavov:2012jq} and Budapest-Wuppertal Collaboration~\cite{Borsanyi:2011sw} for: (a) baryon-strangeness correlations, (b) strangeness fluctuations. Also shown are the HRG results for the discrete mass spectrum (dashed line), continuous mass spectrum obtained by fitting to the PDG (broken-dashed line) and LQCD results (continuous line).}
		\label{fig:fluct}
	\end{figure*}

	In the confined phase of QCD, at finite temperature and density, the effective degrees of freedom are hadrons and their resonances. In the simplest form of the HRG model, the interaction effects and resulting widths of hadrons and resonances are neglected, and the medium constituents are treated effectively as point-like and independent~\cite{BraunMunzinger:2003zd}. Therefore, all the information about the medium composition enter through the mass spectrum. In this contribution, we consider the continuum-extended Hagedorn mass spectrum~\cite{Lo:2015cca}, namely

	\begin{equation}
		\label{eq:def:Hagedorn}
		\rho(m) = \sum_i d_i\delta(m-m_i) + \rho^{H}(m)\theta(m-m_x) \textrm{,}\;\;\;\textrm{where}\;\;\;\rho^{H}(m) = \frac{A\;e^{m/T_H}}{(m^2 + m_0^2)^{5/4}} \textrm.
	\end{equation}

	\noindent The factor $d_i = (2J_i+1)$ is the spin degeneracy of particle $i$ with mass $m_i$, less than the mass of the first resonance $m_x$. $T_H$ is the Hagedorn limiting temperature, $A$ and $m_0$ are additional free model parameters. Noting the mass spectrum in Eq.~(\ref{eq:def:Hagedorn}) is additive, it can be decomposed into a sum of mesonic and baryonic contributions, as well as a sum of particles with definite quantum numbers. Here we assume that the Hagedorn temperature is the same for all different quantum number sectors~\cite{Lo:2015cca}.

	In the grand canonical ensemble, the total thermodynamic pressure of an uncorrelated gas of particles (and antiparticles) is given by

	\begin{equation}
		\label{eq:def:pressure}
		\hat P \equiv \frac{P}{T^4} = \pm \int \dd m\; \rho(m) \int \frac{\dd \hat p}{2\pi^2}\; \hat p^2 \left[ \ln\left(1\pm\lambda e^{-\hat \epsilon}\right) + \ln\left(1 \pm \lambda^{-1}e^{-\hat\epsilon}\right)\right] \textrm,
	\end{equation}

	\noindent where $\hat p = p/T$, $\hat m = m/T$, $\hat \epsilon = \sqrt{\hat p^2 + \hat m^2}$ and $\pm$ sign refers to baryons and mesons respectively. The fugacity term reads $\lambda = \exp\left(B\hat\mu_B + S\hat\mu_S + Q\hat\mu_Q\right)$, where $\hat\mu = \mu/T$. For particles with $B=S=Q=0$, the antiparticle term must be neglected to avoid double counting. Note that the mass spectrum $\rho(m)$ in Eq.~(\ref{eq:def:pressure}) is a model dependent quantity. This allows us to study the effects of different medium compositions.

	In the HRG model, the second order fluctuations of conserved charges are probed by general susceptibilities, i.e., by taking the second derivatives of~Eq.(\ref{eq:def:pressure}), with respect to the chemical potential, namely

	\begin{equation}
		\label{eq:def:fluctuations}
		\hat \chi_{\rm xy} = \frac{\partial^2 \hat P}{\partial \hat \mu_{\rm x} \partial \hat \mu_{\rm y}} \Bigg|_{\hat \mu_x = \hat \mu_y = 0} \textrm,
	\end{equation}

	\noindent where $({\rm x}, {\rm y})$ are conserved charges, which, for our purpose, we restrict to the net-baryon number $B$ and net-strangeness $S$.

	\begin{figure*}[t!]
		\centering\subfigure[]{\includegraphics[width=.45\columnwidth]{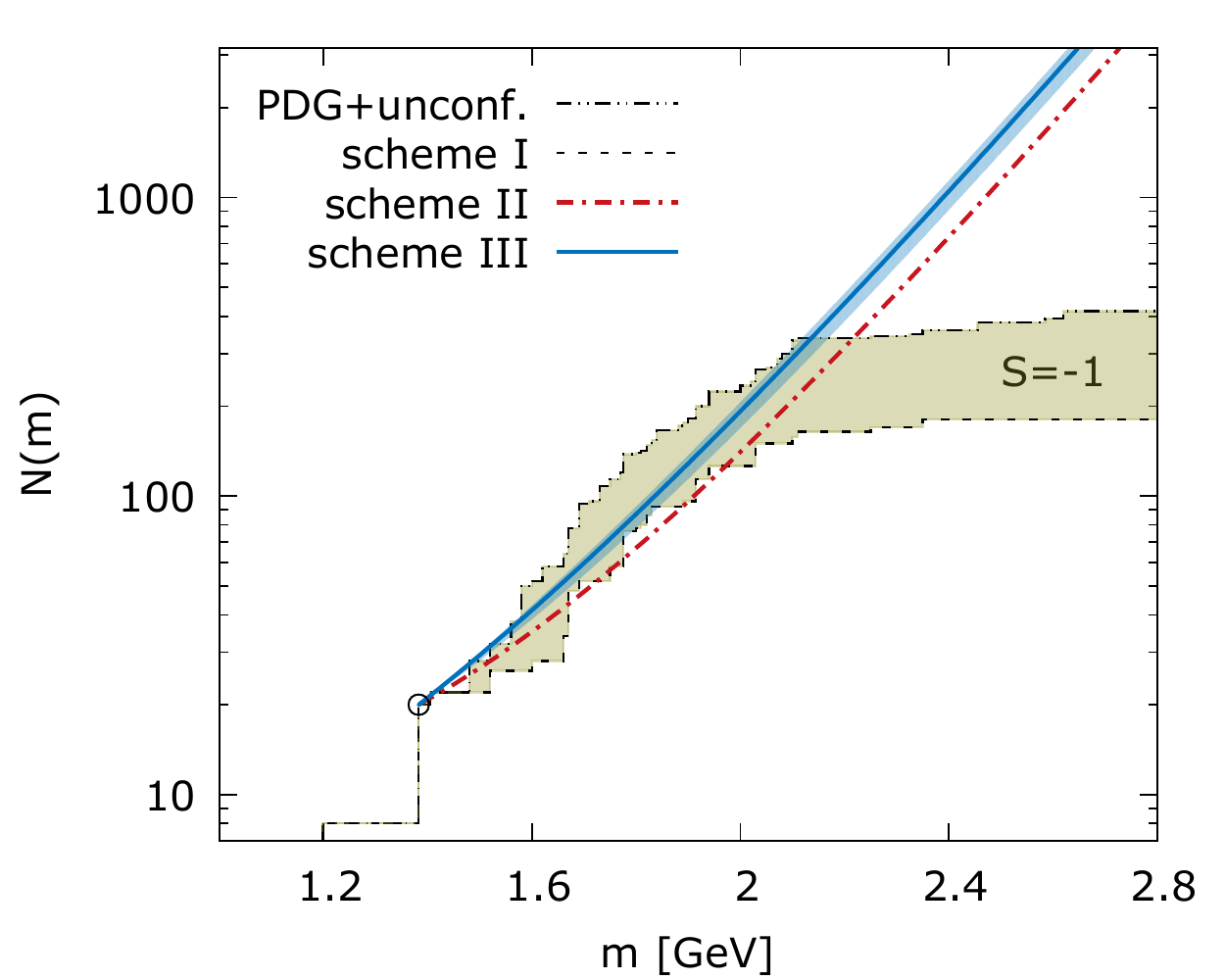}
			\label{fig:cumulant:baryons}}
		\centering\subfigure[]{\includegraphics[width=.45\columnwidth]{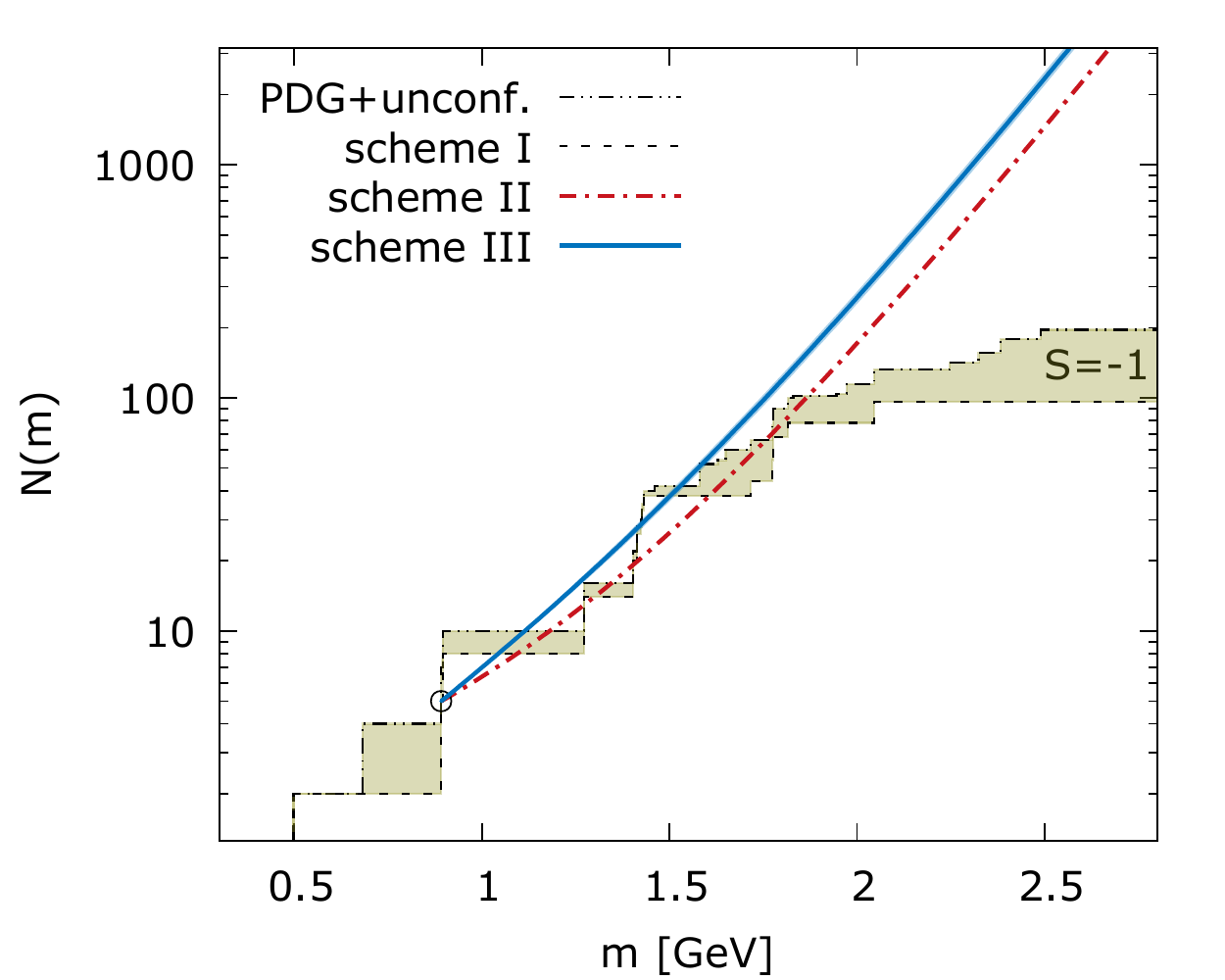}
			\label{fig:cumulant:mesons}}
		\caption{Cumulative PDG mass spectra~\cite{Lo:2015cca} of strangeness $-1$ (dashed line) for baryons (a) and mesons (b). Also shown are the cumulative spectra that include the unconfirmed states (doubly broken-dashed line). The broken-dashed lines and the continuous lines are obtained by fitting the continuous spectra to the PDG and to LQCD results respectively. The errors in the continuous lines come from the fitting procedure in the latter case~\cite{Lo:2015cca}.}
		\label{fig:cumulant}
	\end{figure*}

	To study the implications of different medium compositions on the observables, we consider a comparison of the predictions of the discrete mass spectrum, obtained by neglecting the continuous part of~Eq.(\ref{eq:def:Hagedorn}) (scheme I), with the predictions of the Hagedorn mass spectrum, extracted from the PDG (scheme II) and from the lattice results on $\hat\chi_{\rm BS}$ and $\hat\chi_{\rm SS}$ (scheme III). For scheme I, we take all the established hadrons from the PDG database. Matching the Hagedorn spectrum to the LQCD results in scheme III was performed with the additional assumption that all the missing strange contribution in the baryonic sector comes solely from the $|S|=1$ sector~\cite{Lo:2015cca}.

	In Fig.~\ref{fig:fluct}, we compare obtained results for the baryon-strangeness correlations $\hat\chi_{\rm BS}$ and strangeness fluctuations $\hat\chi_{\rm SS}$. The corresponding cumulative spectra are shown in Fig.~\ref{fig:cumulant}. The discrepancies between the HRG and LQCD results for the baryon-strangeness correlations and strangeness fluctuations are clearly visible in scheme I (Fig.~\ref{fig:fluct}). The spectra extracted in scheme II show that they can be, partially, resolved only at sufficiently high temperatures. This finding already indicates the existence of additional resonances, in the intermediate mass region of the PDG data, that could contribute substantially, due to their lower masses. The matching performed in scheme III shows that the disparities can be completely resolved. Moreover, the corresponding $|S|=1$ baryonic spectrum, shown in Fig.~\ref{fig:cumulant:baryons}, follows the trend of the unconfirmed states in the PDG. However, the $|S|=1$ mesonic spectrum obtained in scheme III, shown in Fig.~\ref{fig:cumulant:mesons}, exceeds that of the PDG, even after the inclusion of the unconfirmed states. This may suggest the existence of still unobserved resonances in the strange-mesonic sector below $2$ GeV.

\section{Conclusions}
\label{section:conclusions}

	Motivated by the recent Lattice QCD (LQCD) results on the baryon-strangeness correlations and strangeness fluctuations, which suggest the existence of unconfirmed resonances in the strange sector of the Hadron Resonance Gas (HRG), we compared the influence of different medium compositions on the HRG results. In particular, we compared the predictions of the experimentally established discrete mass spectrum with the continuous Hagedorn mass spectra, obtained from the fit to the PDG and to the LQCD observables.

	We have shown that the continuous Hagedorn mass spectra, extracted from the PDG data, partially remove the disparities with the lattice results. On the other hand, the Hagedorn mass spectra, matched to LQCD results for fluctuations, not only remove those differences but also point to existence of extra states in the \mbox{strange-hadronic} sector. As a result, the extracted strange-baryonic spectrum is consistent with the trend dictated by the unconfirmed states in the PDG. For the strange-mesonic spectrum, however, it suggests the existence of yet-undiscovered states in the intermediate mass region.

	Nonetheless, one has to remember that, in the HRG model, the contribution to $\hat\chi_{\rm BS}$ and $\hat\chi_{\rm SS}$ comes only from the hadronic sector with open strangeness, whilst, in general, the contribution from the non-strange sector is also possible. 

	Morevoer, the point-like approximation for broad resonances is known to overestimate their contribution to the observables. Hence, they have to be treated with great care when implemented in effective models~\cite{Friman:2015zua, Broniowski:2015oha, Blaschke}. Further investigation and new LQCD results are needed to clarify these issues.

\section*{Acknowledgments}
	\label{section:acknowledgements}

	The author acknowledges the financial support from the organizers of 15th International Conference on Strangeness in Quark Matter and the Helmholtz International Summer School "Dense Matter 2015", fruitful discussions with collaborators P.~M.~Lo, K.~Redlich and C.~Sasaki. The author would also like to thank M.~A.~R.~Kaltenborn for proofreading the manuscript. This work was partly supported by the Polish National Science Center (NCN), under Maestro Grant No. DEC-2013/10/A/ST2/00106.

\end{document}